\documentclass[aps,preprint,showpacs,preprintnumbers,amsmath,amssymb,footi
nbib]{revtex4-1}
\usepackage{amsmath}
\usepackage{graphicx,color,mathrsfs,flafter
}
\usepackage[colorlinks=true, pdfstartview=FitV, linkcolor=red, citecolor=blue, urlcolor=blue]{hyperref}

\begin{document}

\preprint{BNL-99123-2013-JA, RBRC-1000}

\title{Zero interface tension at the deconfining phase transition
for a matrix model of a $SU(\infty)$ gauge theory}
\author{Shu Lin}
\email{slin@quark.phy.bnl.gov}
\affiliation{RIKEN/BNL, Brookhaven National Laboratory, 
Upton, NY 11973}
\author{Robert D. Pisarski}
\email{pisarski@bnl.gov}
\affiliation{
Department of Physics, Brookhaven National Laboratory, 
Upton, NY 11973}
\affiliation{RIKEN/BNL, Brookhaven National Laboratory, 
Upton, NY 11973}
\author{Vladimir V. Skokov}
\email{vskokov@quark.phy.bnl.gov}
\affiliation{
Department of Physics, Brookhaven National Laboratory, 
Upton, NY 11973}
\begin{abstract}
Using a matrix model,
we model the deconfining phase transition at nonzero temperature
for a $SU(N)$ gauge theory at large $N$.
At infinite $N$ the matrix model 
exhibits a Gross-Witten-Wadia transition.  We show that as a consequence,
both the order-disorder and the order-order interface tensions
vanish identically at the critical temperature $T_d$.
We estimate how
these quantities vanish in the matrix model as $T \rightarrow T_d$
and as $N \rightarrow \infty$.  
The numerical solution of the matrix
model suggests possible non-monotonic behavior in $N$ for
relatively small values of $N \sim 5$.
\end{abstract}
\maketitle

\section{Introduction}

Because of dimensional transmutation, the properties of $SU(N)$ gauge
theories are of great interest.  Once one mass scale is set,
all other properties are in principle determined.  For example, if
the theory exhibits a deconfining phase transition at a temperature $T_d$,
then the nature of the phase transition is completely determined once
the value of $T_d$ is known.

Understanding the deconfining phase transition at small $N$ requires numerical
simulations on the lattice
\cite{DeTar:2009ef, Petreczky:2012rq}.  
Using these results, a matrix model was developed to model deconfinement.
These models involve
zero \cite{Meisinger:2001cq, Meisinger:2001fi}, one
\cite{Dumitru:2010mj}, and two 
\cite{Dumitru:2012fw, Kashiwa:2012wa} parameters.
They are soluble analytically for
two and three colors, and numerically for four or more colors.

This matrix model is also soluble in the limit of infinite $N$
\cite{Pisarski:2012bj}.  The phase transition at infinite $N$ is
exceptional, and can be termed a Gross-Witten-Wadia transition
\cite{Gross:1980br, Wadia:1979vk, Wadia:1980cp}.
For the deconfining transition, such a transition was first seen to 
occur on a femto-sphere
\cite{Sundborg:1999ue, Aharony:2003sx, Aharony:2005bq, Schnitzer:2004qt, *AlvarezGaume:2005fv, *AlvarezGaume:2006jg, *Hollowood:2009sy, *Hands:2010zp, *Hollowood:2011ep, *Hollowood:2012nr, Dumitru:2004gd}.  
For such a transition, at infinite $N$ it exhibits aspects of
both first and second order phase transitions.  It is of first
order in that the latent heat is nonzero and proportional to $\sim N^2$.
It is also first order in that the Polyakov loop jumps
from $0$ to $\frac{1}{2}$ at $T_d$. 
On the other hand, it is of second order in that several quantities,
such as the specific heat, exhibit nontrivial critical exponents.
Such an unusual transition only occurs at infinite $N$, as at
finite $N$ the transition is of first order.

Consider a phase transition which can be characterized by the
change of a single field.
If the transition is of first order, 
then at the transition temperature there are two degenerate minima,
with a nonzero barrier between them.  
Then the order-disorder interface tension is nonzero, given by the 
probability to tunnel between the two minima.
Conversely, if the transition is of second order, at the
transition temperature the two minima coincide.
Then there is no barrier to tunnel between
them, and the order-disorder interface tension vanishes.

In a gauge theory asking about the potential is more subtle.
In a matrix model of an $SU(N)$ gauge theory, the potential
exists in $N-1$ dimensions (the number of commuting diagonal generators).
A physical question is to ask how the interface tensions behave,
especially at the phase transition.  There are 
two such interface tensions.  There is the usual order-disorder
interface tension, which exists only at $T_d$.  There are also
order-order interface tensions
\cite{Bhattacharya:1990hk,Bhattacharya:1992qb,Giovannangeli:2002uv,Giovannangeli:2004sg,Hartnoll:2006hr,Armoni:2008yp}
, which are directly related to
the 't Hooft loops for $Z(N)$ charges 
\cite{KorthalsAltes:1999xb,KorthalsAltes:2000gs}.  These exist
for all $T \geq T_d$.  

In this paper we consider the interface tensions 
in the matrix model near $T_d$.  We find that 
in the matrix model, both the order-order
interface tension and the order-disorder interface tensions,
vanish identically at $T = T_d$ when $N = \infty$.  

There is a simple heuristic explanation for our results.  Consider
the potential for the simplest Polyakov loop, that in the fundamental
representation, $\ell_1 = (1/N) {\rm tr}\, {\bold L}$, where $\bold L$ is
the thermal Wilson line.  Then the Gross-Witten-Wadia transition
occurs because at infinite $N$, the potential for $\ell_1$ is completely flat
between $0$ and $\frac{1}{2}$.  That is, at $T_d$ there are two
distinict minima, as expected for a first order transition.  Nevertheless,
because the potential is flat at $N = \infty$
between the two minima, there is no barrier
to tunnel between them, and so the interface tensions vanish.
Such a flat potential was first found for the Gross-Witten-Wadia transition
on a femto-sphere 
\cite{Sundborg:1999ue, Aharony:2003sx, Aharony:2005bq, Schnitzer:2004qt, *AlvarezGaume:2005fv, *AlvarezGaume:2006jg, *Hollowood:2009sy, *Hands:2010zp, *Hollowood:2011ep, *Hollowood:2012nr, Dumitru:2004gd}.  

As noted, in a gauge theory there are other degrees of freedom.  For
example, one can consider higher powers higher powers of $\bold L$,
$\ell_j  = \frac1N {\rm tr}{\bf L}^j$, for $j = 2 \ldots N-1$.  For
the Gross-Witten-Wadia transition, though, all $\ell_j$ vanish at
$T_d$ when $j \geq 2$, which allows us to consider
the potential as a function of a single variable, $\ell_1$.  
We stress, however, that this is true only at $T_d$, and not for
$T \neq T_d$.  Indeed, while we estimate the behavior of the
(order-order) interface tensions for $T > T_d$, we cannot solve the
problem in full generality.  This is because away from $T_d$, {\it all}
$\ell_j$ contribute, and it is a much more difficult problem than
at $T_d$.

The order-disorder interface tension 
has been measured through numerical simulations on the lattice at $N = 3$
by Beinlich, Karsch, and Peikert \cite{Beinlich:1996xg}.
Lucini, Teper, and Wenger give results for 
the order-disorder interface tension for $N = 3$, $4$, and
$6$, and use these to extrapolate to $N = \infty$ \cite{Lucini:2005vg}.
On general grounds the order-disorder interface
tension should be proportional to $\sim N^2 T_d^2$,
with a coefficient which is naturally of order
one.  Instead, Ref. \cite{Lucini:2005vg} finds a very
small coefficient at $N=\infty$, $\approx .014$.  
In the matrix model the coefficient $\sim N^2$ vanishes identically,
and the true behavior is only $\sim N$.  
This is because in the matrix model,
the height of the barrier between the two distinct minima
is small, $\sim 1/N$.  We discuss this further in Sec.
(\ref{sec:finiteN}) and in the Conclusions, Sec. (\ref{sec:con}).

\section{Review of large $N$ thermodynamics}

We are interested in the thermodynamics of pure $SU(N)$ gauge theory 
for temperatures a few times that for the deconfining phase transition
at $T_d$.
The order parameter of the thermodynamics is taken to be the Wilson line:
\begin{align}
{\rm L} (\vec{x})={\rm P} 
\exp \left( ig\int_0^\beta A_0(\vec {x},\tau)d\tau \right).
\end{align}

By a gauge transformation, we can diagonalize the field $A_0$ as
\begin{align}
A_0^{ij}=\frac{2\pi T}{g} \; q_i \; \delta^{ij},
\end{align}
where $i,j=1\cdots N$ and the eigenvalues 
$q_i$ are subject to the $SU(N)$ constraint 
$\sum_i^N q_i=0$. In our model $q_i$ are the fundamental
variables to characterize the transition.
We assume that after integrating out 
the other components of gluon field $A_i$,
that we obtain an effective potential for $q_i$
\cite{Meisinger:2001cq, Meisinger:2001fi,Dumitru:2010mj,Dumitru:2012fw, Kashiwa:2012wa,Pisarski:2012bj}.

\begin{align}\label{mm_V}
&{\tilde V}_{eff}(q)=-d_1(T){\tilde V}_1(q)+d_2(T){\tilde V}_2(q),\\
&
{\tilde V}_n(q)=\sum_{i,j=1}^N|q_i-q_j|^n\left(1-|q_i-q_j|\right)^n.
\end{align}
The potential includes both perturbative ($\tilde V_2$) and non-perturbative ($\tilde V_1$) contributions. The temperature dependent functions $d_1$ and $d_2$ are given by
\begin{align}\label{dn}
d_1(T)=\frac{2\pi}{15}\; c_1\; T^2\; T_d^2,\;
d_2(T)=\frac{2\pi}{3}\left(T^4-c_2T^2T_d^2\right).
\end{align}
 At tree level, the kinetic term is
\begin{align}\label{mm_T}
\tilde {K}(q)=\frac{1}{2} {\rm tr} F_{\mu \nu}^2
=\left(\frac{2\pi T}{g}\right)^2\sum_{i=1}^N\left(\nabla q_i(x)\right)^2.
\end{align}
In a mean-field approximation, the kinetic term does not contribution for a spatially homogeneous
states, and so it can be ignored for 
thermodynamic quantities.
The kinetic term does enter in computing the
interface tension in the following sections.

In the infinite $N$ limit we introduce a continuous variable 
$x=\frac{i}{N}$.  Labeling the eigenvalue $q_i\to q(x)$,
we introduce the eigenvalue density 
\begin{align}
\rho(q)= \lim_{N\to\infty} \frac{1}{N} 
\sum_{i}^N \delta(q-q_i) = \int_{0}^1 dx \delta[q-q(x)] =   \frac{dx}{dq} \; .
\end{align}
At finite $N$, the identities
\begin{align}\label{ids_i}
\sum_i^N1=N,\;\;\sum_i^Nq_i=0
\end{align}
become, at infinite $N$,
\begin{align}\label{ids_rho}
\int dq\rho(q)=1,\;\;\int dq\rho(q)q=0.
\end{align}

The potential is proportional to $N^2$, 
\begin{align}\label{V_N}
&{\tilde V}_n(q)=N^2V_n(q)=N^2\int dx\; dy \; |q(x)-q(y)|^n
\left(1-|q(x)-q(y)|\right)^n \nonumber \\
&=N^2\int dq\; dq'\; \rho(q)\; \rho(q')\; |q-q'|^n\left(1-|q-q'|\right)^n.
\end{align}
This representation transforms the potential into a polynomial in
$q$.  

The minimum of Eq. \eqref{V_N} was  found in Ref.  
\cite{Pisarski:2012bj}.  The solution is
\begin{subequations}
\begin{align}
&\rho(q)=1+b\cos dq,\quad -q_0<q<q_0, \label{rho}\\
& d=\sqrt{\frac{12d_2}{d_1}}. \\
&\cot(d q_0)=\frac{d}{3}\left(\frac{1}{2}-q_0\right)
-\frac{1}{d\left(1/2-q_0\right)} \label{cot},\\
&b^2=\frac{d^4}{9}\left(\frac{1}{2}-q_0\right)^4
+\frac{d^2}{3}\left(\frac{1}{2}-q_0\right)^2+1 \label{bsq}. 
\end{align}
\end{subequations}

For $T>T_d$, $d>2\pi$ and $q_0<\frac{1}{2}$. The eigenvalues do not 
span the full range between $-\frac{1}{2}$ and $\frac{1}{2}$. 
The density is discontinuous at the end points $\rho(\pm q_0)>0$. 
For $T=T_d^+$, $q_0=\frac{1}{2}$ and $d=2\pi$, 
the density is continuous for all values of $q$ in $[-1/2,1/2]$. 
In particulalr it vanishes at the end points $\rho(\pm q_0)=0$. 
For $T<T_d$, the theory is in confined phase, with a uniform
distribution of eigenvalues over the unit circle, 
Eq.~\eqref{rho} with $q_0=\frac{1}{2}$ and $b=0$. 

For the potential Eq. \eqref{V_N} and the eigenvalue distribution 
Eq. \eqref{rho},
at $T_d$, $q_0=\frac{1}{2}$, and the potential is independent of $b$.
Changing $b$ from $0$ to $1$ interpolates between 
confined and deconfined phase, but does not change the potential. 
Hence $b$ is a zero mode of the potential corresponding 
to changing the overall shape of the distribution.
This will play an important role in the construction of the interface. 
For short we call the change of shape related to $b$ the $b$-mode.

It is also worth emphasizing that in the 
derivation of Ref.~\cite{Pisarski:2012bj}, 
we have assumed that the eigenvalue density is symmetric in $q$.
We can obtain different distributions by 
applying an arbitrary $Z_N$ transform to a given solution. 
A $Z_N$ transform of charge $k$, $k = 1 \ldots (N-1)$, is given by
\begin{align}\label{ZN}
q_1, q_2, \cdots q_N\;\;\to q_1+\frac{k}{N}, \cdots 
q_{N-k}+\frac{k}{N}, q_{N-k+1}+\frac{k-N}{N}, \cdots q_N+\frac{k-N}{N}.
\end{align}
Assuming $q_1\le q_2\le\cdots\le q_N$ and $|q_i-q_j|<1$, we can relabel 
the eigenvalues such that they are in an increasing order:
\begin{align}\label{ZN_re}
 q_1, q_2, \cdots q_N\;\;\to q_{N-k+1}+\frac{k-N}{N}, \cdots q_N+\frac{k-N}{N}, q_1+\frac{k}{N}, \cdots q_{N-k}+\frac{k}{N}.
\end{align}

In the infinite $N$ limit, the $Z_N$ tranform takes the following form. 
Define the inverse function of $x(q)=\int_{-q_0}^q dq'\rho(q')$
as $Q(x)$,
\begin{align}\label{O2}
q=Q(x) \to 
q=\left\{\begin{array}{ll}
Q(x+1-\Delta)-1+\Delta& 0<x<\Delta\\
Q(x-\Delta)+\Delta& \Delta<x<1
\end{array}
\right..
\end{align}
Since the potential is invariant under $Z_N$ transformations,
smooth changes in $\Delta$ are another zero mode of the potential,
which we call the shift mode.  
This is also relevant to the construction of the interface. 

We stress that both the $b$ and shift modes are 
become zero modes only at infinite $N$.
The former is because of the
flatness of the potential.  The latter is  
because the $Z_N$ symmetry becomes a continuous $U(1)$ at $N = \infty$.
We comment on what happens at finite $N$ later.

\section{Interface tension}

An interface is a topological object interpolating between two vacua 
of the theory. Suppose the two vacua are separated in the $z$ direction and 
extended in the $x-y$ plane. The effective action has an area law when 
the transverse size $L_{tr}$ is large: $S_{eff}=\alpha L_{tr}$. The 
proportionality constant defines the interface tension. Up 
to cubic order in the perturbative expansion, the 
order-order interface tension 
exhibits Casimir scaling 
\cite{Giovannangeli:2002uv,Giovannangeli:2004sg}
\begin{align}
\alpha\propto k(N-k).
\end{align}
An important question to address is whether the order-order interface
tension satisfies Casimir scaling in the matrix model.

An interface tension necessarily involves a spatial gradient along
the $z$ direction.  Consequently, the kinetic term must be included.  
Since the potential is simple when written in terms 
of the eigenvalue density $\rho(q)$, it is useful
to write the kinetic term in terms of the same variable as well.
Assuming that there is a spatial
gradient only along the $z$ direction, 
\begin{align}
{\tilde K}(q)=N^2\; K(q)
=N^2\frac{(2\pi T)^2}{g^2N}\int dzdx
\left(\frac{\partial q(x,z)}{\partial z}\right)^2,
\end{align}
where the partial derivative is taken at fixed $x$. 

We start with the eigenvalue density 
$\rho(q,z)=\partial x(q,z)/\partial q$.
Assuming that the range of the eigenvalue distribution does not change
over the interface, 
\begin{align}
x(q,z)=\int_{-q0}^q dq'\rho(q',z).\end{align}
Using the chain rule, we have
\begin{align}\label{dqdz}
&\frac{\partial q(x,z)}{\partial z}=-\frac{\partial q(x,z)}{\partial x}\frac{\partial x(q,z)}{\partial z} \nonumber\\
&=-\frac{\partial_z\int^q dq'\rho(q',z)}{\rho(q,z)}.
\end{align}
The kinetic term becomes (c.f. \cite{Polchinski:1991tw})
\begin{align}\label{T_N}
K(q)=\int dzdq\frac{\left(\partial_z\int^q dq'\rho(q',z)\right)^2}{\rho(q,z)}.
\end{align}

\subsection{Interface tension at $T_d$}

For simplicity consider the interface tension at $T_d$ first. 
At the transition there are both order-order interface and order-disorder 
interfaces. Due to the complicated form of the kinetic energy Eq. \eqref{T_N}, 
solving for the full solution of the interface seems to be hopeless. 
However, the presence of the zero modes allows us to show 
that both interface tensions vanish. 
It is straightforward to construct an interface using the shift mode:
\begin{align}\label{shift}
q=\left\{\begin{array}{ll}
Q(x+1-\Delta f(z))-1+\Delta f(z),& 0<x<\Delta f(z);\\
Q(x-\Delta f(z))+\Delta f(z),& \Delta f(z)<x<1
\end{array}
\right.
\end{align}
with $f(-L)=0$ and $f(L)=1$ at  two boundaries of the interface. A $k$-wall interpolating two vacua related by $Z_N^k$ transformation corresponds to $\Delta=\frac{k}{N}$.

The idea is to take $f(z)=\frac{z}{2L}+\frac{1}{2}$ such that 
$\frac{\partial q}{\partial z}\sim\frac{1}{L}$. 
As $f$ is a zero mode, this not change the potential energy, 
while the kinetic energy is supressed by $\frac{1}{L}$. 
In the limit $L\to\infty$, action vanishes as $1/L$,
and so the interface tension vanishes at $T_d$.

This does not work for the shift mode. To see that, 
we need to take a close look at the kinetic term. 
On the interface, the density of eigenvalues is given by
\begin{align}
\rho=\left\{\begin{array}{ll}
1+\cos2\pi(q+1-\Delta f(z)),& Q(1-\Delta f(z))-1+\Delta f<q<-1/2+\Delta f(z)\\
1+\cos2\pi(q-\Delta f(z)),& -1/2+\Delta f(z)<q<Q(1-\Delta f(z))+\Delta f(z).
\end{array}
\right.
\end{align}
Integrating with respect to $q$, we obtain
\begin{align}
x=q+\frac{1}{2}+\frac{1}{2\pi}\sin [2\pi(q-\Delta f)],
\quad Q(1-\Delta f)-1+\Delta f<q<Q(1-\Delta f)+\Delta f,
\end{align}
from which it follows
\begin{align}\label{gradient_f}
\frac{\partial q}{\partial z}=-
\frac{\cos [2\pi(q-\Delta f)]\Delta f'}{1+\cos [2\pi(q-\Delta f)]}.
\end{align}
Plugging the gradient Eq. \eqref{gradient_f} into Eq. \eqref{T_N}, we 
identify a non-integrable singularity at $q=1/2+\Delta f$. 
Therefore, we conclude that the interface built by the shift mode 
is ruled out by the divergent kinetic energy.

A second possibility is to build an interface with the $b$-mode. The two 
vacua at the ends of the interface are joined through a confining phase 
in the middle. Defining $Q_b$ as the inverse function of
\begin{align}
x=q+ \frac{1}{2}+\frac{b}{2\pi}\sin (2\pi q).
\end{align}
At $T=T_d$, $Q_1(x)$ reduces to the distribution for the deconfined phase,
and $Q_0(x)$ for the confined phase.
The interface is constructed as 
\begin{align}\label{b_interface}
\text{part I}: &q=Q_{b(z)}(x),\; -L<z<0 \leftrightarrow \nonumber\\
\text{part II}: &q=\left\{\begin{array}{ll}
Q_{b(z)}(x+1-\Delta)-1+\Delta,& 0<x<\Delta;\\
Q_{b(z)}(x-\Delta)+\Delta,& \Delta<x<1.
\end{array}
\right.\; 0<z<L,
\end{align}
with the boundary conditions $b(\pm L)=1$, $b(0)=0$. The distributions 
from part I and part II joining at $z=0$ are identical and given explicitly by $q=x-1/2$. 
We note that part II is a $Z(N)$ transform of part I, flipped in $z$.
We will show below that the kinetic energy is not divergent,
so that we can apply the previous argument to arrive at the suppression
in $1/L$. 
Since the path given by the first and second lines of 
Eq. \eqref{b_interface} are 
related by $Z_N$ transformations, they necessarily have the same 
kinetic and potential energies. 
It is sufficient to restrict ourselves to the first line. 
It is easy to find the gradient
\begin{align}\label{gradient_b}
\frac{\partial q}{\partial z}=-\frac{b'}{2\pi}
\frac{\sin2\pi q}{1+b\cos 2\pi q}, 
\end{align}
where here and in the following primes 
denote the derivatives with respect to the argument. 
We see possible singularities at $q=\pm1/2$ 
from the denominator are cancelled by the numerator, 
giving rise to a finite result for the kinetic energy. 
For a given $b$, it is not difficult to compute the 
integral in $q$ by contour integration,
\begin{equation}
\label{b_kin}
\int_{-1/2}^{1/2} dq\rho(q)
\left(\frac{\partial q}{\partial z}\right)^2=\left(\frac{b'}{2\pi}\right)^2
\frac{-b^4+b^2\left(5-3\sqrt{1-b^2}\right)+
4\left(-1+\sqrt{1-b^2}\right)}{b^2\sqrt{1-b^2}\left(-1+\sqrt{1-b^2}\right)^2}.
\end{equation}
We checked that the kinetic term is finite in the limit $b\to1$.

\subsection{Interface tension at $T>T_d$}

The order-disorder interface tension is only defined at $T_d$.
Above $T_d$, the order-order interface tension, equivalent to
the 't Hooft loop, is nonzero.
However, we were not able to compute the order-order interface tension
in full generality. 

Let's consider two limiting situations, 
$\Delta\to0$ and $T\to T_d$. We consdier them in turn.

In the case $\Delta\to0$, the end points of the interface are given by
\begin{align}\label{quote}
q=Q(x),\;\;q=\left\{\begin{array}{ll}
Q(x+1-\Delta)-1+\Delta,& 0<x<\Delta;\\
Q(x-\Delta)+\Delta,& \Delta<x<1.
\end{array}
\right.
\end{align}
The eigenvalues between $\Delta<x<1$ have change infintesimally, 
while those between $0<x<\Delta$ have a finite jump;
the latter, however, are suppressed because there are few of them.
With this in mind, we write down the following path for the interface:
\begin{align}\label{Delta0}
&q=\left(Q(x+1-\Delta)-1+\Delta-Q(x)\right)g(z)+Q(x),& 0<x<\Delta; \\
&q=\left(Q(x-\Delta)+\Delta-Q(x)\right)f(z)+Q(x),& \Delta<x<1.
\end{align}
The unknown functions $g(z)$ and $f(z)$ interpolate between
$0$ and $1$. They could in principle depend
on $x$, which characterizes the change of shape of the eigenvalue density. To
leading order in $\Delta$ we can ignore this dependence. 
Now we can work out the potential energy along the path
\begin{align}
V=\int_0^\Delta dxdy V\left(q(x)-q(y)\right)
+\int_\Delta^1 dxdy V(q(x)-q(y))+2\int_0^\Delta dx
\int_\Delta^1 dy V(q(y)-q(x)).
\end{align}
The first term is of order $O(\Delta^3)$ and may be ignored. The second 
term starts with the vacuum potential energy as that at leading order. 
The third term is of order $O(\Delta)$. We need to know the 
$O(\Delta)$ correction of the potential to the vacuum one:
$$
\delta V=\int_0^1dxdy
\frac{\partial V}{\partial \lvert q(x)-q(y)\rvert}
\Delta\lvert Q'(x)-Q'(y)\rvert f
$$
\begin{equation}
+2\int_0^\Delta dx\int_0^1dy \left(V(q(y)+q_0)-V(q(y)+q_0+(1-2q_0)g)\right).
\label{dV}
\end{equation}
Evaluating Eq. \eqref{dV} for the vacuum solution Eq. \eqref{rho}, 
we find the first term vanishes identically, while the second term gives
\begin{align}
\delta V=\frac{d^2}{12}(1-2q_0)^4g^2(1-g)^2.
\end{align}
The kinetic energy reads 
\begin{align}
&\int dzdx\left(\frac{\partial q}{\partial z}\right)^2 \\
\text{with}\; &\frac{\partial q}{\partial z}=\left\{\begin{array}{ll}
(2q_0-1)g',& 0<x<\Delta;\\
\Delta\left(1-Q'(x)\right)f',& \Delta<x<1.
\end{array}
\right.
\end{align}
We ignore the contribution from $\Delta<x<1$ 
because it is of order $O(\Delta^2)$. 
Combining the kinetic and potential terms, we have
\begin{align}
\int dz(K+\delta V)=\int dz\left(\Delta (1-2q_0)^2g'^2+\Delta\frac{d^2}{12}(1-2q_0)^4g^2(1-g)^2\right).
\end{align}
We need to minimize the above action with the boundary condition 
$g(-\infty)=0$ and $g(\infty)=1$. Using a trick
(if $(x-y)^2\ge0$, $x^2+y^2\ge 2xy$), we can 
obtain the minimum without solving for $g$:
\begin{align}
\int dz(K+\delta V)&\ge \Delta\int dz 
2\sqrt{(1-2q_0)^2g'^2\frac{d^2}{12}(1-2q_0)^4g^2(1-g)^2} \nonumber\\
&=\Delta\int dg g(1-g)\frac{d(1-2q_0)^3}{\sqrt{3}}=
\Delta\frac{d(1-2q_0)^3}{6\sqrt{3}},
\end{align}
which leads to
\begin{align}
\alpha=\Delta\frac{d}{6\sqrt{3}}\left(1-2q_0\right)^3.
\end{align}

Now we look at the other limiting case: $T\to T_d$. 
We expect the interface  to mimic the $T=T_d$ case, {\it i.e.} 
two vacua joined through a confined distribution. 
We consider the following simple path:
\begin{align}\label{simple}
\text{part I}:&\rho=a(z)+b(z)b_0\cos dq,\;-q_0<q<q_0, \nonumber\\
\text{part II}:&\rho=\left\{\begin{array}{ll}
\frac{1}{2q_0},& -q_0-(1-\Delta)(1-2q_0)f(z)<q<-(1-2\Delta)q_0-(1-\Delta)(1-2q_0)f(z);\\
\frac{1}{2q_0},& -(1-2\Delta)q_0+\Delta(1-2q_0)g(z)<q<q_0+\Delta(1-2q_0)g(z),
\end{array}\right., \nonumber\\
\text{part III}:&\rho=\left\{\begin{array}{ll}
a(z)+b(z)\cos d(q+1-\Delta),& Q_b(1-\Delta)-1+\Delta<q<q_0-1+\Delta; \\
a(z)+b(z)\cos d(q+1-\Delta),& -q_0+\Delta<q<Q_b(1-\Delta)+\Delta,
\end{array}\right.,
\end{align}
We have $-L<z<-\frac{L}{2}$ for part I, $-\frac{L}{2}<z<\frac{L}{2}$ for part II and $\frac{L}{2}<z<L$
 for part III, with the following boundary conditions
 \begin{align}
b(\pm L)=1,\;b(\pm L/2)=0,\; f(-L/2)=0,\;f(L/2)=1.
\end{align}
In Eq.\eqref{simple} $q_0$ and $b_0$ are determined by Eqs. \eqref{cot} and Eq. \eqref{bsq}, 
respectively. 
The normalization condition $\int dq\rho(q)=1$ 
forces $a(z)q_0+\frac{b(z)}{d}\sin dq_0=1/2$,
while the tracelessness condition $\int dq\rho(q) q=0$ 
sets $f(z)=g(z)$. Furthermore, we have the boundary conditions
The interface is composed of three parts. Part 
I makes use of the $b$-mode to deform the eigenvalue distribution. 
At $T\to T_d$, $b$ mode is nearly a zero mode, 
with minimum cost of potential energy. 
Part II is unique to the case $T>T_d$, 
with the motion of the eigenvalues suppressed 
by $(1-2q_0)$. Part II is just a $Z_N$ transform of part I, flipped in $z$.

Let us look at part I first. 
The potential energy with respect to the vacuum one 
has the following expansion in $\frac{1}{2}-q_0$
(which is an effective expansion in $T-T_d$)
\begin{align}
\delta V=\frac{60-d^2}{90}(1-b)^2
\left(\frac{1}{2}-q_0\right)^2+O\left(\left(\frac{1}{2}-q_0\right)^3\right).
\end{align}
The kinetic energy can be taken as at $T=T_d$, 
ignoring higher order corrections Eq. \eqref{b_kin}. 
With these ingredients, we can already work out 
the contribution to the interface tension from part I. 
It is worth noting prior doing any computations that
$\delta V\sim \left(\frac{1}{2}-q_0\right)^2$, $K\sim O(1)$.

Next we consider part II. This case is particular 
easy because of the constant 
eigenvalue density in this part. 
To leading order, the potential energy and kinetic energy are
\begin{align}\label{patch}
&\delta V=\frac{1}{90}\bigg[60-d^2-60f(1-f)
\Delta(1-\Delta)\left(12-d^2\Delta(1-\Delta)\right)\bigg]
\left(\frac{1}{2}-q_0\right)^2, \nonumber\\
&T=f'^2\Delta(1-\Delta)\left(\frac{1}{2}-q_0\right)^2.
\end{align}
We note in part II both potential and kinetic terms are suppressed by
$\left(\frac{1}{2}-q_0\right)^2$, leading only to contributions
of higher order in $(T-T_d)$ for the interface tension.

Adding up contributions from all three parts, 
we obtain a contribution to leading order which is twice that of part I. 
It is not difficult to convince ourselves that 
$\alpha\sim\left(\frac{1}{2}-q_0\right)$ and it is  independent of $\Delta$.

We will improve the result by considering 
a more sophisticated ansatz. Note that in Eq. \eqref{simple}, 
we have chosen to turn on the $b$-mode and 
separating $f$-mode separately. 
Here we consider more general ansatz by turning them on simultaneously. 
Furthermore, as we learn from the previous example 
that part II has only subleading contributions, 
we consider the abovementioned modification to part I only. 
With these in mind, we consider the following ansatz
\begin{align}
&\rho=a(z)+b(z)\cos d(q+(1-\Delta)f_L(z)),
\;-q_0-(1-\Delta)f_L<q<Q_b(\Delta)-(1-\Delta)f_L
\nonumber\\
&\rho=a(z)+b(z)\cos d(q-\Delta f_R(z)),\;
Q_b(\Delta)+\Delta f_R<q<q_0+\Delta f_R.
\end{align}
The tracelessness condition forces $f_L=f_R\equiv f$. 
It is also natural to require $f\sim 1-2q_0$ 
such that the eigenvalues do not overseparate along the path. 
The introduction of the $f$-mode induce corrections 
to the potential in Eq.~\eqref{patch}. 
Defining $f={\bar f}\left(\frac{1}{2}-q_0\right)+O\left((\frac{1}{2}-q_0)^2\right)$, 
we only need to consider corrections up to the second order in ${\bar f}$. 
It turns out the first order in ${\bar f}$ has a coefficient 
$P(\Delta,b)(\frac{1}{2}-q_0)+O\left((\frac{1}{2}-q_0)^2\right)$ 
and the ${\bar f}^2$ term has a coefficient 
$Q(\Delta,b)+O\left((\frac{1}{2}-q_0)\right)$. 
When both are taken into account, we have for the potential
\begin{figure}
\includegraphics[width=0.4\textwidth]{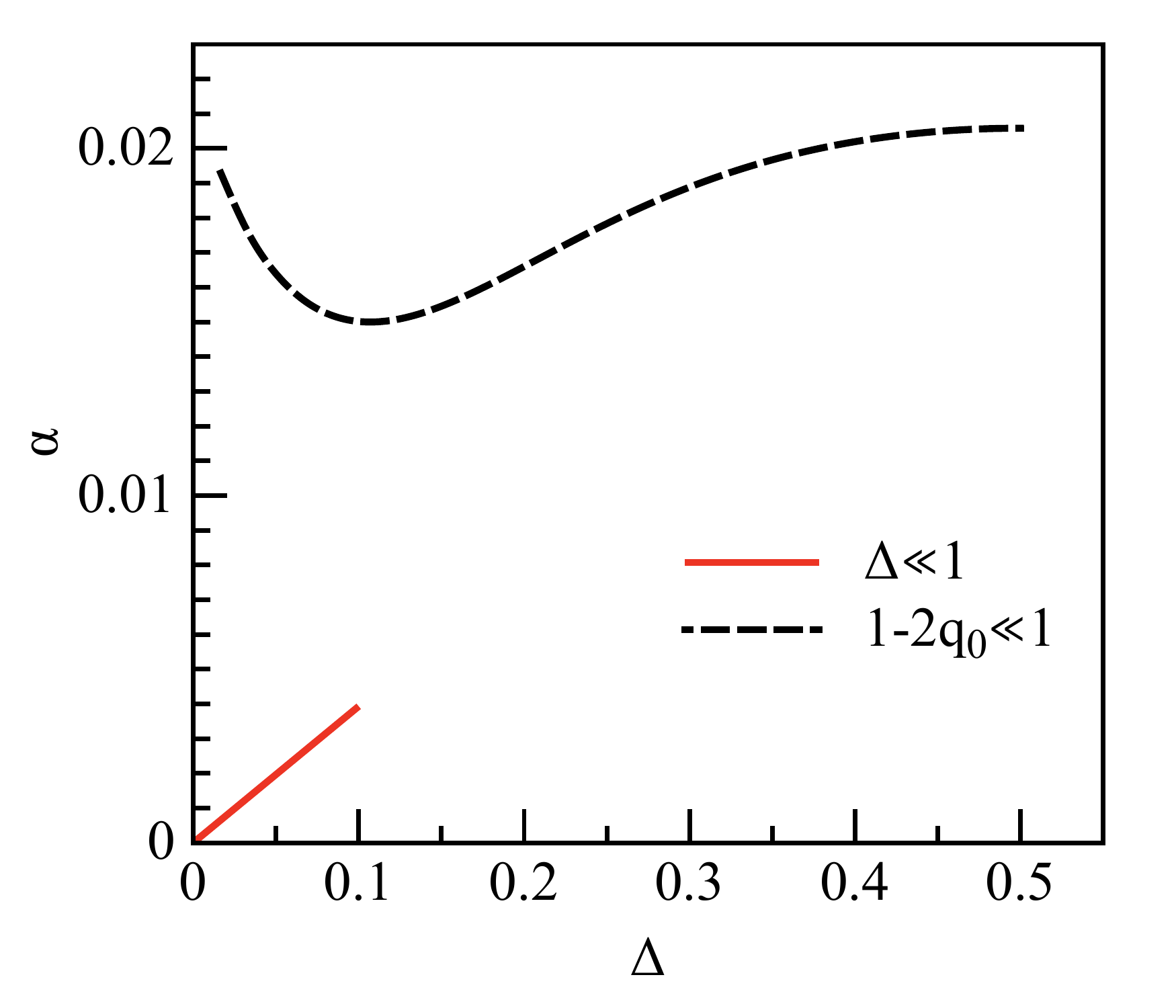}
\caption{Interface tension at $q_0=0.3$. 
Due to the symmetry $\Delta\leftrightarrow 1-\Delta$, 
we have only shown half range of $\Delta$. 
The red linear line is for  $\Delta\ll1$,
while the black dashed line is obtained assuming $1-2q_0 \ll 1$ (i.e. $T \to T_d$).}\label{fig_interface}
\end{figure}
\begin{align}
\delta V=\left(\delta V_0(b)+P(\Delta,b){\bar f}+Q(\Delta,b){\bar f}^2\right)\left(\frac{1}{2}-q_0\right)^2,
\end{align}
where
\begin{align}
&\delta V_0=\frac{1}{90}\bigg[60-d^2-60f(1-f)\Delta(1-\Delta)\left(12-d^2\Delta(1-\Delta)\right)\bigg], \\
&P=-\frac{1}{48}(b-1)\left(32b+(1-4Q_b(\Delta)^2)(-48+d^2(1+8Q_b(\Delta)+12Q_b(\Delta)^2-16\Delta Q_b(\Delta)))\right) \nonumber\\
&-16b(-1+12Q_b(\Delta)^2)\cos dQ_b(\Delta), \\
&Q=384b^2-48bd^2\left(1+4Q_b(\Delta)^2-Q_b(\Delta)(4-8\Delta)\right)+d^2\bigg[48-192Q_b(\Delta)^2 \nonumber\\
&+d^2\left(-5-24Q_b(\Delta)^2+48Q_b(\Delta)^4+Q_b(\Delta)^3(32-64\Delta)+16\Delta(1-\Delta)+24Q_b(\Delta)(-1+2\Delta)\right)\bigg] \nonumber\\
&+48b\left(8b+d^2(1-4Q_b(\Delta)^2)\right)\cos dQ_b(\Delta).
\end{align}
We have also factored out the overall 
$\left(\frac{1}{2}-q_0\right)^2$ dependence. 
One important property we confirm numerically is that $Q>0$. 
To lower the potential energy, we choose ${\bar f}-\frac{P}{2Q}$. As a result,
\begin{align}
\delta V=\left(\delta V_0-\frac{P^2}{4Q}\right)\left(\frac{1}{2}-q_0\right)^2.
\end{align}
At the same time, the correction also introduces 
$\Delta$ dependence to the interface tension. 
With the same kinetic energy as at leading order, 
we obtain the interface tension 
\begin{align}
\alpha=2 \int dz \sqrt{\delta V K}\sim (\frac{1}{2}-q_0).
\end{align}
The final result is obtained numerically 
and shown in Fig. (\ref{fig_interface}).  Both scenarios are included.

We close this section with the observation that 
these different scenarios do not have a common region in which
they are both valid.
This suggests the limits $T\to T_d$ and $\Delta\to0$ (or $N\to\infty$) 
do not commute. Therefore, it 
is important to evaluate corrections at finite $N$ near $T_d$.

\section{Finite $N$ correction near $T_d$}
\label{sec:finiteN}

Corrections at finite $N$ enter where the integrals at $N = \infty$
are replaced by discrete sums.
This can be evaluated with Euler-MacLaurin formula \cite{Goldschmidt:1979hq}
\begin{align}\label{EM}
\sum_{i=1}^NF(i)=N\int_{\frac{1}{N}}^1dxf(x)+
\frac{f(1)+f(\frac{1}{N})}{2}+
\sum_{k=1}^\infty\frac{B_{2k}}{(2k)!N^{2k-1}}
\left(f^{(2k-1)}{}'(1)-f^{(2k-1)}{}'\left(\frac{1}{N}\right)\right),
\end{align}
where $f(i/N)=F(i)$ and $B_{2k}$ are Bernoulli numbers. 
We wish to shift the argument and the lower integration 
bound from $\frac{1}{N}$ to $0$. This reshuffle can be done with a Taylor 
expansion. We end up with the following surprisingly simple expression:
\begin{align}\label{EM2}
\sum_{i=1}^NF(i)=N\int_{0}^1dxf(x)+\frac{f(1)-f(0)}{2}+
\sum_{k=1}^\infty\frac{B_{2k}}{(2k)!N^{2k-1}}
\left(f^{(2k-1)}{}'(1)-f^{(2k-1)}{}'(0)\right).
\end{align}
We are not able able to prove Eq. \eqref{EM2}, 
which requires a recursion relation among Bernoulli numbers. 
We have verified explicitly that it holds 
for the first few terms in the expansion.
Applying Eq. \eqref{EM2} to the sums, we have
\begin{align}\label{sums}
N&=\sum_i^N1=N\int_0^1 dx \\
0&=\sum_i^Nq=N\int_0^1 dx q+\frac{q(1)-q(0)}{2}+\cdots \\
V&=\sum_{i,j=1}^NV(q_i,q_j)=
N^2\int_0^1dxdyV(q(x)-q(y))+
2N\int_0^1 dx\frac{V(q(x)-q(1))-V(q(x)-q(0))}{2} \nonumber\\
&+\cdots,
\end{align}
where $\cdots$ denote terms higher order in $\frac{1}{N}$ expansion. 
As before, we define the eigenvalue density $\rho(q)=\frac{dx}{dq}$, 
in terms of which 
Eq. \eqref{sums} becomes
\begin{align}\label{sums_rho}
1&=\int_{q_-}^{q_+} dq\rho(q) \\
0&=\int_{q_-}^{q_+} dq\rho(q)q+\frac{q_+-q_-}{2N}+\cdots \\
V&=N^2\int_{q_-}^{q^+}dqdq'\rho(q)\rho(q')V(q-q')
+N\int_{q_-}^{q_+}dq\rho(q)\left(V(q_+-q)-V(q-q_-)\right)+\cdots,
\end{align}
with $q_-=q(0)$ and $q_+=q(1)$. For the infinite $N$ eigenvalue distribution, 
$\rho(q)=\rho(-q)$ and $q(0)+q(1)=0$. 
The correction at next to leading order to $V$ vanishes identically. 
However this need not be true for the corrected distribution in 
$\frac{1}{N}$.

To determine the new eigenvalue distribution, we vary 
$V$ with respect to $\rho$, 
subject to the usual normalization and tracelessness constraints in 
Eq. \eqref{sums_rho}. This gives 
\begin{subequations}
\begin{align}
&2\int_{q_-}^{q_+}dq'\rho(q')V(q-q')+\frac{1}{N}\left(V(q_+-q)-V(q-q_-)\right)+\lambda_1+\lambda_2q=0 \label{var_rho} \\
&\int_{q_-}^{q_+} dq\rho(q)=1 \label{norm} \\
&\int_{q_-}^{q_+} dq\rho(q)q+\frac{q_+-q_-}{2N}=0 \label{trace}.
\end{align}
\end{subequations}
We follow the method of \cite{Pisarski:2012bj} in solving for $\rho$. Taking derivative with respect to $q$ four times, we arrive at
\begin{align}
\rho''(q)+d^2(\rho(q)-1)=0.
\end{align}
The general solution is given by
\begin{align}\label{rho_cs}
\rho(q)=1+b\cos dq+c\sin dq.
\end{align}
Plugging it into Eq. \eqref{var_rho}, we find 
the result is organized as a fourth order polynomial in $q$. The 
coefficients of $q$ and the constant term can always be set to zero by choice of the Lagrange multipliers $\lambda_1$ and $\lambda_2$. The 
remain coefficients of $q^4$, $q^3$ and $q^2$ give 
three independent equations. Together with Eq. \eqref{norm} 
and Eq. \eqref{trace}, we have in total five equations, to 
be satisfied by four constants $b$, $c$ and $q_{\pm}$. 
It turns out when four of the equations are satisfied, 
the fifth automatically holds. In practice, 
we solve for $\cos dq_{\pm}$ and $\sin dq_{\pm}$ 
in terms of the constants $b$, $c$ and $q_{\pm}$. 
Defining $q_s=q_-+q_+$ and $q_d=q_+-q_-$, 
we can first find an equation for $q_d$
\begin{align}\label{qd}
\cot\frac{dq_d}{2}=-\frac{12-d^2(1-q_d)^2}{6d(1-q_d)}.
\end{align}
It is easy to see $q_d=2q_0$ as defined in Eq. \eqref{cot} is 
free of corrections in $1/N$, and that 
\begin{align}\label{trig}
&q_s=\frac{1}{N} \; \frac{-12-2d^2q_0(1-2q_0)^2}{d^2(1-2q_0)^2}\\
&\frac{\left(144+12d^2(1-2q_0)^2+d^4(1-2q_0)^4\right)\left(36+d^2N^2(1-2q_0)^4\right)}{d^2(1-2q_0)^4}=144(b^2+c^2)N^2
\end{align}
The first shows that $q_s$ is suppressed in $1/N$.  The second combined with
\begin{align}
\tan\frac{dq_s}{2}=\frac{1-\cos d(q_-+q_+)}{\sin d(q_-+q_+)},
\end{align}
can be used to determine $b$ and $c$. 
To have a consistent $\frac{1}{N}$ expansion, we need to have
\begin{align}\label{bc}
b=b_0+\frac{b_2}{N^2}+\cdots,\;c=\frac{c_1}{N}+\cdots.
\end{align}
At leading order $b_0$ agrees with Eq. \eqref{bsq}. 
$c_1$ is
\begin{align}\label{c1}
c_1=-b_0 \; d\; q_0.
\end{align}

We have obtained Eqs. \eqref{rho_cs}, \eqref{qd}, \eqref{trig}, 
\eqref{bc} and \eqref{c1} as the new eigenvalue distribution up
to order $\frac{1}{N}$. The appearance of $\sin(d q)$ in Eq. 
\eqref{rho_cs} looks worrisome at first sight: the 
Polyakov loop can have a nonvanishing imaginary part, 
violating charge conjugation symmetry. The cause of the 
strange behavior lies in our $\frac{1}{N}$ expansion. 
We have taken Eq. \eqref{sums} literally as the 
$\frac{1}{N}$ expanison and used it to solve for the 
eigenvalue distribution $\rho$ to the first nontrivial order. 
However, the result contains constants which organize 
themselves in $\frac{1}{N}$ expanison, 
making our naive expansion inconsistent. 
A consistent expansion has to take 
into account the $\frac{1}{N}$ factors in the constants.

Using the experience from before, 
we expect that to order $\frac{1}{N}$, the constants have the following form:
\begin{align}\label{constants}
&b=b_0,\;c=\frac{c_1}{N},\; q_d=2q_0,\; q_s=\frac{q_{s1}}{N}.
\end{align}
Plugging Eq. \eqref{constants} into Eqs. \eqref{var_rho}, 
\eqref{norm}, \eqref{trace} and keeping terms up to order $\frac{1}{N}$, 
we find, surprisingly, that the resulting equations can be solved by
\begin{align}
c_1=0,\;q_{s1}=-\frac{12}{d^2(1-2q_0)^2}.
\end{align}
\begin{figure}
\includegraphics[scale=0.3]{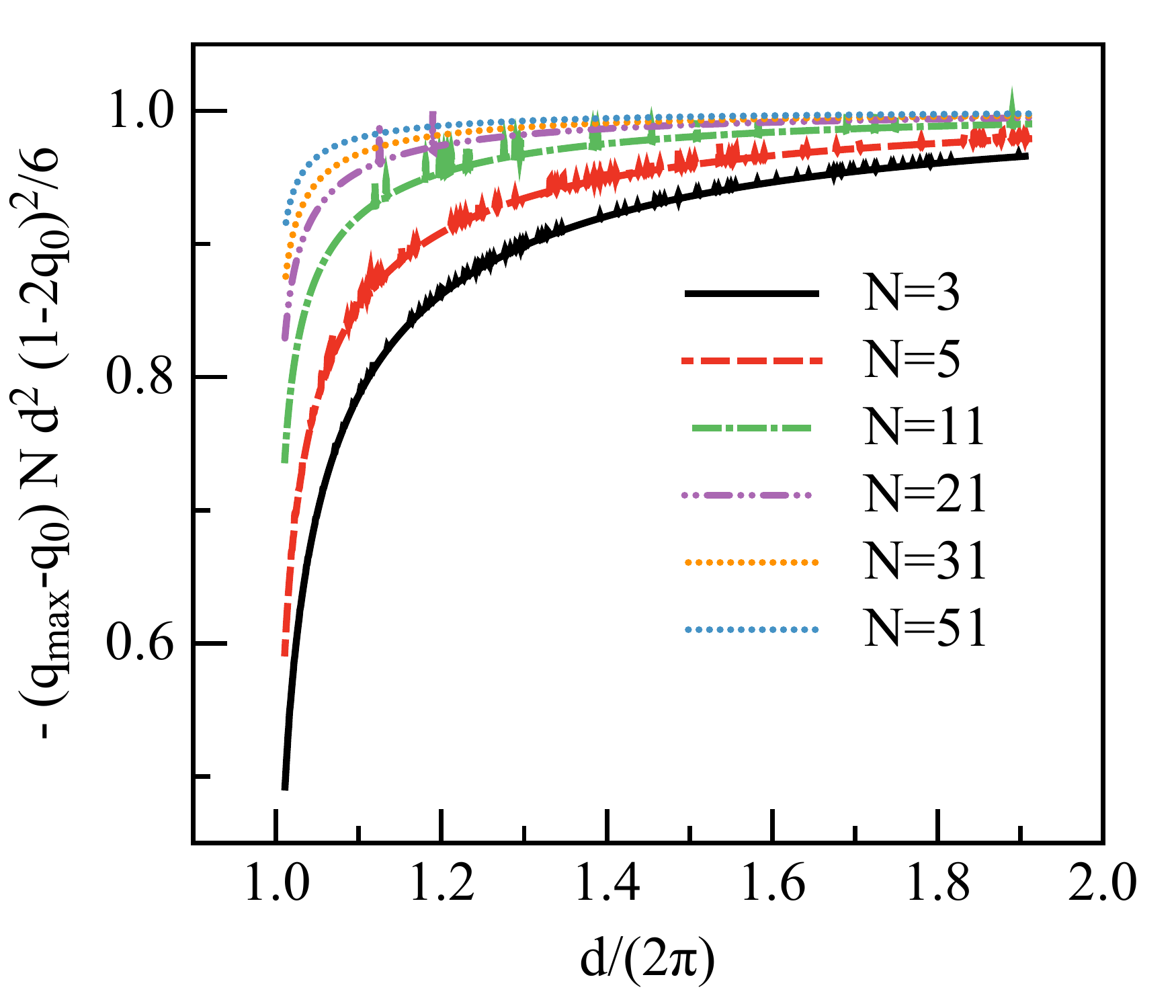}
\caption{
The maximal eigenvalue, $q_{\rm max}$, as a function of 
$d$ for different number of colors, $N$.}
\label{fig:qmax}
\end{figure}
The term $\sin(d \, q)$ naturally drops out, and we find
explicitly that the imaginary part of all Polyakov loops
vanish to order $\frac{1}{N}$.
We summarize the eigenvalue distribution as follows
\begin{align}\label{distr}
&\rho=1+b\cos dq,\;q_-<q<q_+\\
&q_-+q_+=-\frac{12}{d^2N(1-2q_0)^2},\;q_+-q_-=2q_0.
\end{align}
Thus we have
\begin{align}\label{qplus}
& -(q_+-q_0) d^2 N (1-2q_0)^2/6 =1.
\end{align}
In Fig. (\ref{fig:qmax}), 
we compare this combination with the 
numerical results for the maximal eigenvalue, $q_{\rm max}$, 
which at high $N$ is approximated by $q_+$, 
for different number of colors and different $d$, 
that is temperature. Nice agreement between analytic expression
Eq. \eqref{qplus} and numerical results is seen. From 
Eq. \eqref{distr} and our numerical simulations we also see that
when $N(1-2q_0)^2\sim1$, the large $N$ expansion breaks down.
We numerically computed that, at $d=2\pi$, i.e. $T=T_d$, we have $q_{max}-q_0\sim1/\sqrt{N}$,
in contrast to the behaviour of the maximal eigenvalue in the confined and 
the deconfined phase $q_{max}-q_0\sim1/N$. 

\begin{figure}
\includegraphics[scale=0.3]{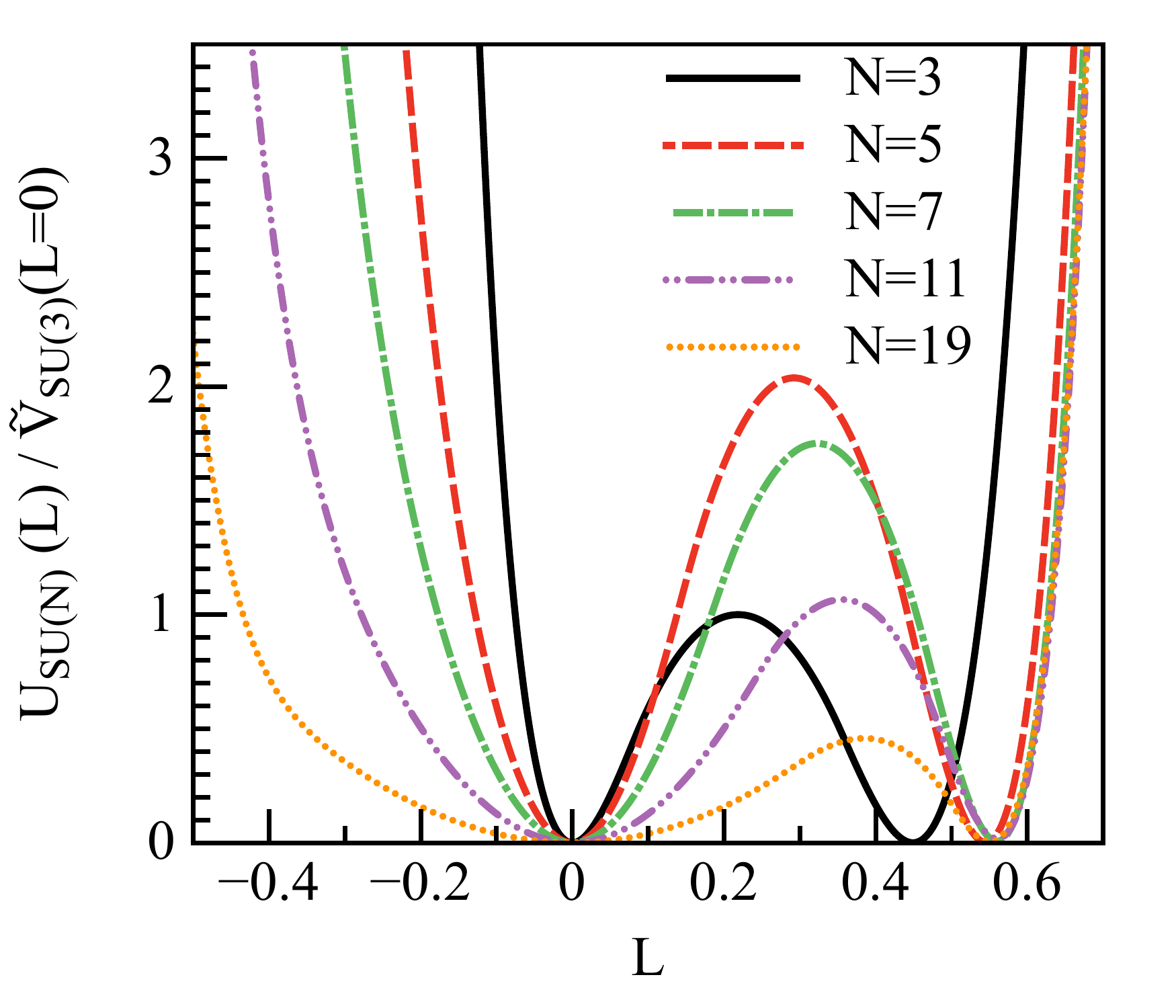}
\caption{
The non-equilibrium potential $U(L) = \tilde{V}(L) -  \tilde{V}(L=0) $.  
}
\label{fig:Ul}
\end{figure}
\begin{figure}
\includegraphics[scale=0.3]{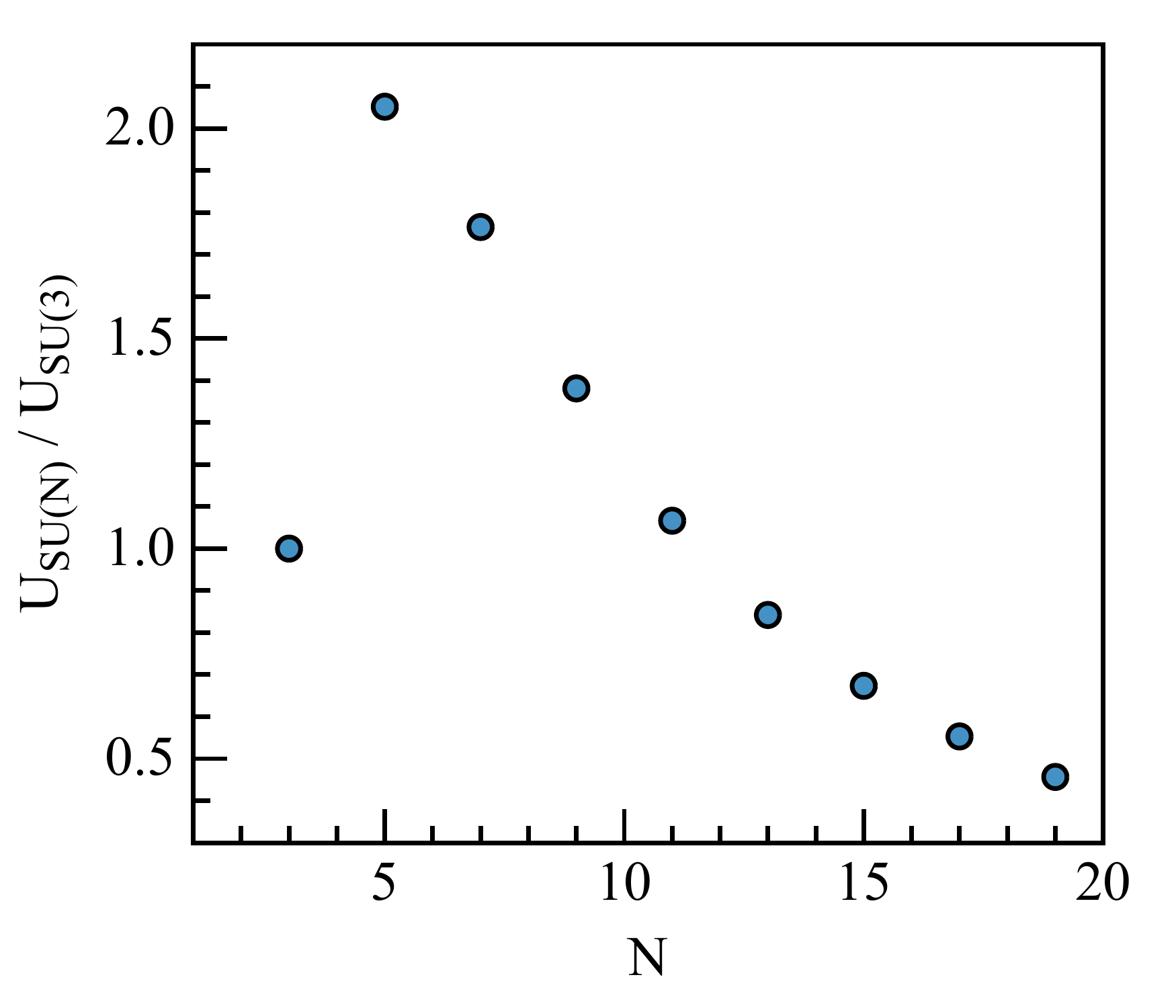}
\caption{
The maximum of the potential $U_{\rm max}$ for different $N$. 
}
\label{fig:Umax}
\end{figure}

With the correct eigenvalue distribution, 
we can proceed to evaluate the correction to the potential. 
We find the correction vanishes to order $\frac{1}{N}$. 
To obtain the correction to the next order, 
we need to find $\rho$ at the corresponding order. 
We will not do it at this time, but use Eq.
\eqref{distr} to give an estimate of the correction. 
We need the following formula for the potential
\begin{align}
V&=\int_{q_-}^{q_+}dqdq'\rho(q)\rho(q')V(q-q')+\frac{1}{N}\int_{q_-}^{q_+}dq\rho(q)\left(V(q_+-q)-V(q-q_-)\right) \nonumber\\
&-\frac{V(q_+-q_-)}{2N^2}+\frac{B_2}{N^2}\int_{q_-}^{q_+}dq\rho(q)\left(\frac{V'(q_+-q)}{\rho(q_+)}+\frac{V'(q-q_-)}{\rho(q_-)}\right)+\cdots.
\end{align}
Taking into account $\frac{1}{N}$ correction to $q_s$, we have $\frac{1}{N^2}$ contribution from all three naive orders. We will not spell out the detail of the calculation, but only list the final results:
\begin{align}
&V_1=\frac{180-5d^4q0^2(1-2q_0)^5+24d^2q_0^2(-5+20q_0-30q_0^2+16q_0^3)}{30d^2N^2(1-2q_0)^3} \\
&V_2=-\frac{q_0(1-2q_0)(6-d^2q_0(1-2q_0))}{3N^2} \\
&V_3=\frac{q_0(1-2q_0)(6-d^2q_0(1-2q_0))}{6N^2},
\end{align}
where $V_1$, $V_2$ and $V_3$ denote contributions from naive order $1$, $\frac{1}{N}$ and $\frac{1}{N^2}$ respectively. We note the large $N$ expansion breaks down as $N^2(1-2q_0)^3\sim1$. The bulk thermodynamic quantity has a less stringent criterion than the eigenvalue distribution.

We were able to analytically estimate corrections to the potential
at the minumum only. Numerically it is possible to go beyond this and 
compute the potential as a function of the Polyakov loop $L$. In 
Fig. (\ref{fig:Ul}), 
the potential normalized by $N^2-1$ is shown as a function of $L$ 
for different 
 number of colors.  
It is remarkable, that the potential flattens from the side of the confinement point $L_d=0$. 
The potential maximum characterizes the order-disorder and order-order interface tensions and 
it falls like $1/N^2$  for large N, which is in qualitatively agreement
with large $N$ expansion, as demonstrated in Fig. (\ref{fig:Umax}).

\vskip 1cm

\section{Conclusions}
\label{sec:con}

The matrix model studied here is clearly only one of many possible
matrix models.  Its advantage is that it can be solved analytically
at infinite $N$, and numerically at finite $N$.
In this model all interface tensions vanish at the 
deconfining phase transition.  Since 
the transition is of first order,
this would be striking evidence that it is an unusual transition,
perhaps of the Gross-Witten-Wadia type.

Presently, numerical simulations of $SU(N)$ gauge theories can only be
carried out at relatively small $N$, $N < 10$.  
For two colors the order-disorder interface tension vanishes, as
the transition is of second order.  
For three colors 
one the order-disorder interface tension is relatively
small \cite{Beinlich:1996xg}.  This presumably reflects that
the transition for three colors is weakly first order, because of
its proximity to the second order transition for two colors.
This leads one to expect that as $N$ increases, that the order-disorder
interface tension, divided by $N^2$, 
increases monotonically from $N=3$, and becomes constant at infinite
$N$.

The numerical solution of the
matrix model indicates the contrary, that the order-disorder
interface tension, divided by $N^2$, behaves non-monotically with $N$.  From 
Fig. (\ref{fig:Umax}), the barrier of the potential, suitably normalized,
increases from $N=3$ to $N=5$, and then slowly decreases as $N$ increases
further.

Such non-monotonic behavior in $N$ is unexpected, and could well
just be an artifact of the model.  This could be settled
by numerical studies on the lattice of the order-disorder interface
tension for moderate values of $N \sim 5$.  It might even
provide hints of a Gross-Witten-Wadia transition at infinite $N$.

\begin{acknowledgments}
The research of R.D.P. and V.S. is supported
by the U.S. Department of Energy under contract \#DE-AC02-98CH10886.
S.L. is supported by the RIKEN Special Postdoctoral Researchers Program.
\end{acknowledgments}


%

\end{document}